\documentclass[onecolumn,11pt]{elsarticle}
\usepackage{epstopdf}
\usepackage{epsfig}
\usepackage{amssymb}
\usepackage{color}
\usepackage{bm}
\usepackage{amsfonts}
\usepackage{lineno,hyperref}
\usepackage{array}
\usepackage{microtype}
\usepackage[a4paper, total={6.25in, 8.5in}]{geometry}

\begin{document}

\begin{frontmatter}

\title{Lorentzian wormholes supported by tachyon matter}

\author{ Rikpratik Sengupta$^a$, Shounak Ghosh$^b$, Mehedi Kalam$^a\footnote{$^*$Corresponding author.\\
{\it E-mail addresses:} rikpratik.sengupta@gmail.com, (RS), shounakphysics@gmail.com (SG), kalam@associates.iucaa.in (MK).}$}

\address{$^a$Department of Physics, Aliah University, Kolkata 700 160, West Bengal, India\\
$^b$ Directorate of Legal Metrology, Department of Consumer Affairs, Govt. of West Bengal, Alipurduar 736121, West Bengal, India}






\begin{abstract}
Wormholes with Ellis geometry have been successfully constructed using tachyon matter~\cite{Das}. However, for such a wormhole, it is obtained that the redshift function is necessarily a constant, and also the wormhole is plagued with an imaginary tachyon potential and a constant field if the solutions are obtained in the absence of a cosmological constant term. So, a physically plausible wormhole solution is possible only in the presence of a $\Lambda$ term. In this paper, we try to construct a wormhole from tachyon matter with three $other$ geometries \textit{different} from the Ellis geometry and see whether it is possible to construct them successfully, besides checking whether the restrictions of the Ellis wormhole can be overcome with these geometries. Among others, we obtain one very interesting result that for all three of these geometries \textit{different} from the Ellis, the $\Lambda$ term is no longer an essential ingredient in constructing physically plausible traversable wormholes and the tachyon matter, capable of providing explanations for the ``\textbf{dark sector}" of the universe,is itself sufficient for this purpose.
\end{abstract}

\maketitle
\end{frontmatter}

\section{Introduction}

In 1962 Fuller and Wheeler~\cite{FH1962} first coined the term `wormhole' to define a tunnel-like connection
between two different points of spacetime in the same universe or two different universes. The wormhole is physically
represented by a set of the mathematical solution obtained by solving the Einstein Field Equations (EFEs) found by Einstein
himself along with Rosen~\cite{ER1935}. In this study, we have dealt with a particular kind of traversable Lorentzian
wormhole supported by exotic tachyon matter. Wormholes obtained from General Relativity (GR), residing in a Lorentzian
manifold are known as Lorentzian wormholes. However, they often appear to be unstable in the absence of exotic matter
at the throat violating the null energy condition (NEC) in GR~\cite{MT1988}. Macroscopic wormholes of Lorentzian signature
may act as traversable shortcuts in spacetime, allowing the intriguing possibility of time-travel, thus acting
as potential time-machines~\cite{MTY,FN,V}. Fuller and Wheeler~\cite{FH1962} argued that the solutions obtained from the
EFE were unstable as the wormhole would be pinched-off at the throat and thus any signal would be trapped in the infinite
curvature region which makes the wormhole non-traversable.

In the year 1988, Morris and Thorne~\cite{MT1988} proposed the idea of traversable wormholes as a set of solutions to the EFE they had studied,
forming a tool for a simpler understanding of Einstein's GR. They showed that a stable wormhole must violate
the NEC of GR so that the pinching-off condition of the throat can be prevented. According
to their prescription for the spacetime metric describing a traversable wormhole, the metric potentials viz. the
redshift and the shape functions need to follow certain conditions: (i) the shape function at the throat radius
must be equal to itself, (ii) the ratio between the shape function and radial parameter $(r)$ must be less than $1$
for the radial separation greater than the throat radius in order for the metric coefficient to be regular, and
(iii) most importantly for the wormhole to be traversable, the flaring-out condition
at the throat must hold good, which simplifies to the condition that the slope of the shape
function at the throat must be less than unity. As we shall discuss later, this condition
effectively describes the violation of the NEC. Finally, for a wormhole to be traversable,
the formation of any horizon must be avoided and this can be achieved via the global finiteness
of the redshift function.

The idea of exotic tachyon matter originates from the instability and decay of objects called $D$-branes
arising in superstring theories~\cite{Sen1,Sen2}. In the effective low energy limit, the potential is found to
have an unstable maximum corresponding to zero tachyon field. Sen~\cite{Sen3} showed that this is justified by
the fact that tachyons exhibit the property $(mass)^2<0$. There is a phenomenon of rolling of the tachyon from
the unstable maximum to a stable average, expected value in the vacuum. This phenomenon finds applications in
cosmology~\cite{Kim,Sami,Paul,Sengupta}. However, our case is different as in the cosmological applications
time-dependent configurations of the tachyon field find use but here we will make use of spatially dependent
configurations~\cite{Liu,Kim2,Gorini}. Wormhole supported by exotic tachyon matter for Ellis geometry has been
studied by Das and Kar~\cite{Das}. Wormhole with other forms of exotic matter at the throat has been studied
in~\cite{Barcelo,Hayward,Picon,Sushkov,Lobo,Zaslavskii,Chakraborty}. \textbf{Thin shell wormholes in heterotic string theory have been investigated in\cite{R1}. A comparative analysis of thin shell wormhole construction in relativistic context and in the context of Horava-Lifshitz modified gravity has been carried out in\cite{R2}. Wormholes have also been constructed using Chaplygin gas\cite{R3} which is a fluid used in cosmology to explain the accelerating universe in the present epoch. Recently wormhole solutions have been obtained in the halo region of different galaxies which gives rise to the possibility of detecting a wormhole\cite{MK}. }

It is well known today that the ``\textbf{dark sector}" of the universe comprising of dark matter and dark energy is responsible for constituting almost $96 \ \%$ of the present universe. Tachyons being superluminal and having characteristic imaginary mass, have been used to explain distinctive cosmological effects, by possibly contributing to a fraction of the dark matter composition~\cite{Davies}. The contribution of tachyons to the dark matter profile may be verified by a modification in the temperature-time relationship of the cosmic fluid besides an increase in particle creation in the early universe due to quantum effects. The tachyon field has also been used to account for the late-time cosmic acceleration~\cite{Srivastava,Bagla,Albrow,Avelino}. The dark energy has been considered to be tachyon driven and coupled non-minimally to the curvature, with the tachyon field described by a self interacting inverse-cubic potential. There is found to be a late-time phase transition from tachyonic dark energy to tachyonic cold dark matter (CDM), suggesting a possible solution to the ``cosmic coincidence" problem~\cite{Srivastava}. Bagla et al.~\cite{Bagla} in their seminal paper considered tachyon matter along with non-relativistic matter and radiation. The density of tachyon was found comparable to the matter density even in the matter-dominated era. Interestingly, for an exponential tachyonic potential, the late-time acceleration phase preceded a matter dominated phase, thus avoiding the possibility of appearance of future singularities from the $\Lambda$-CDM models. Tachyons have also been considered as quanta of dark energy~\cite{Albrow}. A comparative study of the tachyonic and quintessential dark energy models have been done by Avelino et al.~\cite{Avelino}, who have obtained identical evolution of the Hubble parameter and the potential for both the fields, but a striking dissimilarity is highlighted in the evolution of both fields as a function of redshift.  

Although we have taken an exotic matter source, it is essential to check the validity of the NEC as the energy density
and pressure components depend on the tachyon field and the potential whose solution we obtain. So a violation of the NEC
can justify that we have obtained the correct form of the field and potential corresponding to different wormhole geometry.
We generalize the idea of~\cite{Das} to three other geometries characterized by properly defined shape functions different
from the Ellis geometry~\cite{Ellis} and investigate the possibility of constructing wormholes corresponding to these
geometries and their stability. Our present work is organized as follows: in the following section i.e Sec. 2, we have
discussed the mathematical models for the construction of a wormhole and also discussed the mathematical formulations
of the Tachyon fields. In the next sections, i.e., Sec. 2, Sec. 3, and Sec. 4 we have studied the wormhole in three
different geometries defined by three properly defined shape functions respectively. In each section, we have checked
their stability and also tested other essential features in order to investigate the physical viability of the obtained
solutions in each geometry. In Sec. 5 we discuss the results that we obtain and put concluding remarks.

\section{Mathematical models of the wormhole}
In this section we will discuss the mathematical formulation for the construction of a wormhole. In order do so we consider
the general form a static, spherically symmetric line element in terms of the proper radial distance ($l$), describing a stationary
wormhole, can be written as
\begin{equation}\label{eq1}
	ds^2=-\psi^2(l) dt^2+dl^2+r^2(l)(d\theta^2+sin^2\theta d\phi^2).
\end{equation}

Here the proper radial distance ($l$) extends up to infinity both in the positive and negative directions just like the time
coordinate $t$. In the above equation, the functions $\psi(l)$ and $r(l)$ are very essential in describing the wormhole, the
former being known as the redshift function and the later is referred to as the shape function of the wormhole.

The Einstein field equations (EFE) for the line element (1), in the absence of a cosmological constant and with tachyon matter
as the source, takes the form

\begin{eqnarray}\label{eq2}
	-{\frac{2r''(l)}{r(l)}}-{\frac { \left(r'(l) \right) ^{2}}{ (r(l)) ^{2}}}+ \frac{1}{(r(l))^{2}}=-V(T)\sqrt {1+ ({T'(l)})^{2}},\label{eq3}\\
	{\frac {2 \psi'(l) r'(l) }{\psi (l) r(l) }}+{\frac { \left( r'(l) \right) ^{2}}{ (r(l))^{2}}}- \frac{1}{(r(l))^{2}}={\frac {V(T) }{\sqrt {1+ \left( {T'(l)}\right)^{2}}}}, \label{eq4}\\
	{\frac {\psi''(l)}{\psi (l) }}+{\frac {\psi'(l) r'(l) }{\psi(l) r(l) }}+{\frac {r''(l) }{r(l)}}=V(T) \sqrt {1+ \left( {T'(l)}\right)^{2}}.
\end{eqnarray}

Here $T(l)$ denotes the tachyon field, $V(T)$ is the potential of the tachyon field and `prime' denotes derivative with
respect to the proper radial distance $l$. The pressure is anisotropic for tachyon matter with the transverse component
being the exact negative of the energy density. We will follow the prescription by ~\cite{Das}, where we will consider
different shape functions for the wormhole like they considered the Ellis shape function, and obtain the tachyon field
and potential.

Just like it was found in their analysis, we also find that the structure of the differential Eqs. (2)-(4) is such that,
for a redshift function depending on the proper radial distance, irrespective of the choice of the shape function, the
quantity $T'^2$ turns out to be negative, thus implying that the tachyon field is physically unrealistic. We can broadly
generalize their statement as they found it for a particular solution of the redshift function depending on $l$ and also
a particular shape function, but we claim that for any acceptable shape function, the redshift function must be a constant
for obtaining a physically realistic tachyon field. A constant redshift function means $\psi'(l)=0$. For simplicity, we assume
the constant to be unity.

\subsection{Conservation equation and the Tachyon field}

The conservation equation for a fluid with anisotropic pressure has the form
\begin{equation}\label{eq5}
	p_r'+(p_r+\rho)\psi'+\frac{2}{l}(p_t-p_r)=0.
\end{equation}

For a constant redshift function, this reduces to

\begin{equation}\label{eq6}
	{\frac{d p_r}{dl}} +\frac{2}{l}(p_t -p_r )=0.
\end{equation}

The above equation can be written in a simplified form as

\begin{equation}\label{eq7}
	2 \left( {T'(l)} \right) ^{3}-lT''(l) +2T'(l) =0.
\end{equation}

On solving the above differential equation, the solution for the tachyon field has the form
\begin{equation}\label{eq8}
	T(l)=-C_1^{\frac{1}{4}} \left( {\it EllipticF} \left( {\frac {l}{\sqrt [4]{{C_1}}}},I \right) -{\it EllipticE} \left( {\frac {l}{\sqrt [4]{{C_1}}}},I \right) \right),
\end{equation}
where $EllipticF$ and $EllipticE$ represent the incomplete elliptic integral of the first kind and second kind respectively and $C_1$ is an integration constant. The variation of the tachyon field along the proper radial distance has been plotted in Fig. 1.

The tachyon field is obtained to be independent of the shape function. This is an interesting feature as even if we choose a different shape function, the tachyon field will remain the same for a constant redshift function $\psi'(l)=0$.

\begin{figure*}[thbp] \label{TF}
	\centering
	\includegraphics[width=0.5\textwidth]{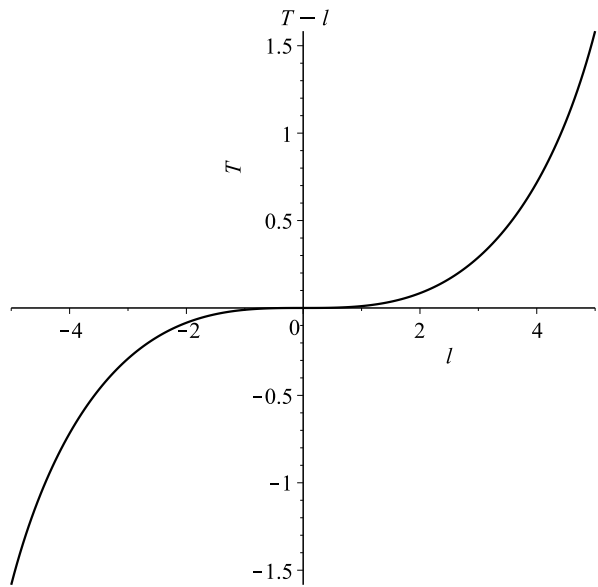}
	\caption{Variation of the T with respect to l.}
\end{figure*}

\section{First shape function}

In this section we make an attempt to construct a traversable wormhole from a well accepted shape function.~\cite{Chakraborty}.
Here we have choosen the form of shape function as follows
\begin{equation}\label{eq9}
	r(l) =r_0 \left(1+{\frac{l}{r_0}} \right)^{n},
\end{equation}
where $r_0$ is the throat radius and $n$ is a parameter which must lie in between 0 and 1. This shape function satisfied
all the essential criteria to be a physically valid shape function as mentioned earlier.

Now plugging in the expression of the tachyon field Eq. (8) and the form of the shape function Eq. (9) in the first EFE of Eq. (2) gives us

\begin{equation}\label{eq10}
	V(l)={\frac{\sqrt{\left(-4{r_0}^{2} \left( n-\frac{1}{2} \right)n R \right)^{2} +3 \left( \left( n-\frac{2}{3} \right){r_0}^{4}{n}^{3} \left( {\frac {R}{r_0}} \right)^{4n}+ R^{4} \right) \left( {\frac {R}{r_0}} \right)^{2n}}}{\left({\frac{R}{r_0}}\right)^{3n} R^{2}{r_0}^{2}}},
\end{equation}
where $(r_0+l)=R$.

For the Ellis wormhole it was found that in the absence of a \textbf{cosmological constant} $\Lambda$, the tachyon potential turned
out to be imaginary and physically unacceptable but here we don't require a $\Lambda$ term to obtain physically acceptable
potential for the tachyon field.

One can notice that $l$ cannot be expressed as a function of $T$ because of its complexity, as a result we are unable to express
$V$ in terms of the tachyon field, but obtain it in terms of the proper radial distance. So, a 3-dimensional plot of the
tachyon potential, field and the proper radial distance has been presented in Fig. 2.

\begin{figure*}[thbp] \label{VLT1}
	\centering
	\includegraphics[width=0.5\textwidth]{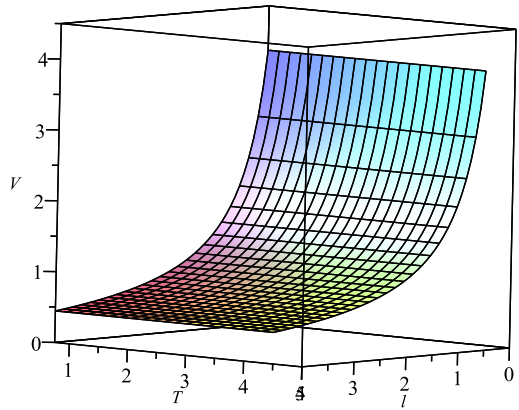}
	\caption{Variation of the V, T with respect to l.}
\end{figure*}

\subsection{Stability}

The linear stability of the metric potential $r$ can be tested under small perturbations. The finite nature
of the linear perturbation in $r$ can be used as a tool to confirm the stability. We assume the perturbations
$\delta r$ to be of much lesser order than $r$.

Owing to the structure of stress-energy tensor $T_{\mu \nu}$ for the tachyon field, the $tt$ and $\theta\theta$
components of the Einstein tensor are equal and opposite in magnitude. This fact gives rise to a differential
equation connecting the redshift and shape functions and their derivatives with respect to the proper radial distance as
\begin{equation}\label{eq11}
	\frac{r''}{r}+\frac{r'^2}{r^2}-\frac{1}{r^2}-\frac{r'\psi'}{r\psi}-\frac{\psi''}{\psi}=0.
\end{equation}

\begin{figure*}[thbp] \label{Rad1}
	\centering
	\includegraphics[width=0.5\textwidth]{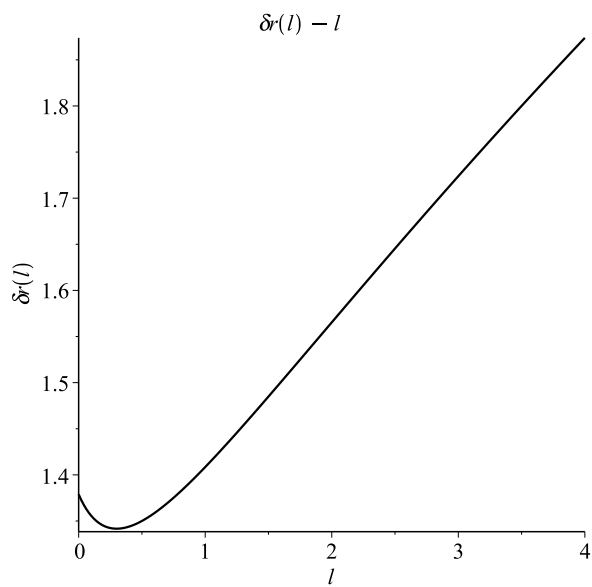}
	\caption{Variation of the perturbation in metric potential with respect to $l$.}
\end{figure*}

As claimed earlier, for obtaining a realistic tachyon field we considered unit redshift function.
Using this, the above equation can be written in a perturbed form as

\begin{equation}\label{eq12}
	\delta'' r(l) +{\frac {2\left(r'(l) \right)\delta' r(l) }{r(l) }}+{\frac{\left( r''(l) \right) \delta r(l) }{r(l) }}=0.
\end{equation}

For our particular form of the shape function, the equation gives a solution of the form

\begin{equation}\label{eq13}
	\delta r(l) ={C_2} (R )^{-n}+{C_3} R^{-n+1},
\end{equation}
where $C_2$ and $C_3$ are the constants of integration.
The perturbations are found to be finite everywhere. Therefore, we confirm stability against linear small
perturbation. A plot of the perturbation in $r$ along the proper radial distance $l$ is presented in Fig. 3.

It is not possible to obtain the perturbation in the tachyon field explicitly as the integration is not possible
for the above differential equation. So we cannot comment on the field perturbation being finite for all $l$.
However as we can see from the expression and plot for the perturbation in the metric potential, it is finite
for all $l$ and so we confirm the stability.

\subsubsection{NEC:}

Among the essential criteria listed earlier for the metric potentials to be a acceptable wormhole solution, one flare-out
condition for the shape function was mentioned. This flare-out condition at the throat is physically responsible
for holding back the throat of the wormhole from collapsing and is essential for its traversability. The condition
imposes a restriction on the stress-energy tensor for the matter inside the throat of the form $T_{\mu \nu}k^\mu k^\nu<0$,
which basically means the matter inside the throat must be of an exotic nature as it has to violate the NEC in GR.
In a nutshell, for successful formation of the wormhole NEC must be violated. The condition for violation of NEC for
anisotropic fluid is $\rho+p_r<0$. From the EFE we get

\begin{equation}\label{eq14}
	\rho+p_r -{\frac {V({T'(l)})^{2}}{\sqrt {1+ ({T'(l)})^{2}}}}.
\end{equation}

\begin{figure*}[thbp] \label{NEC1}
	\centering
	\includegraphics[width=0.5\textwidth]{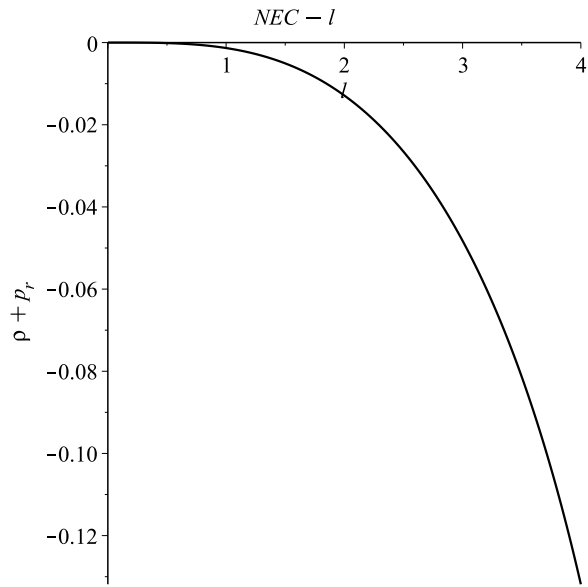}
	\caption{Variation of the NEC with respect to $l$.}
\end{figure*}

Plugging in the solutions for $T(l)$ and $V$ in the above equation

\begin{equation}\label{eq15}
	\rho+p_r=-{\frac {l^4 \sqrt {-2 R^2 (2n-1) {r_0}^{2}n \left( {\frac{R}{r_0}}\right) ^{-2n}+ R^{4} \left( {\frac {R}{r_0}} \right) ^{-4n}+{r_0}^{4}{n}^{3} \left( 3n-2 \right) }}{\sqrt {{C_1} \left( -l^4+{C_1} \right) } R^{2}{r_0}^{2}}}.
\end{equation}
Here $r_0+l=R$. This shows that the NEC is violated for wormhole supported by tachyon matter. A plot for the
violation of NEC along the proper radial distance is given in Fig. 4.

\section{Second shape function}

We consider another well accepted shape function~\cite{Rahaman2} which satisfies all the properties required to describe a
wormhole. The shape function has the form

\begin{equation}\label{eq16}
	r(l) = r_0(1+l)^{\frac{1}{\omega}},
\end{equation}
where $\omega>1$ is a constant parameter.

Using the expression for the tachyon field and shape function in the EFE will yield the potential for the tachyon field as
\begin{equation}\label{eq17}
	V (l) ={\frac{\sqrt{(L)^{\frac{4}{\omega}}\left( 2{\omega}^{2}{r_0}^{2}(L)^{2}(\omega-2) (L)^{\frac{2}{\omega}}-2 \left( \omega-\frac{3}{2} \right) {r_0}^4 (L)^{\frac{4}{\omega}}+{\omega}^{4} (L)^{4} \right) }}{(L)^{2}{r_0}^{2}{\omega}^{2}}\left((L)^{\frac{4}{\omega}} \right) ^{-1}}.
\end{equation}
Here $L=1+l$.
\begin{figure*}[thbp] \label{VLT2}
	\centering
	\includegraphics[width=0.5\textwidth]{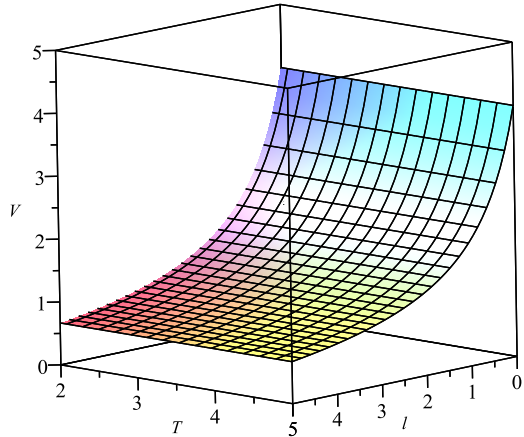}
	\caption{Variation of the V, T with respect to l.}
\end{figure*}

As we can see the potential obtained is real and does not require a $\Lambda$ term to be inserted in the
EFE to obtain physically realistic potential as in the case of an Ellis wormhole. However, as the tachyon
field is obtained from the conservation equation to be independent of the shape function and has a form
that cannot be separated from the proper radial distance $l$, so we cannot express the potential as a
function of the tachyon field, but we obtain it as a function of $l$. So we plot a three dimensional graph
for showing the variation between $V$, $T$ and $l$ in Fig. 5.

\subsubsection{Stability}

Here also we will test the linear stability of the metric potential under small perturbations. The perturbation
$\delta r \ll r$. As stated earlier, even for this shape function, $T'^2$ becomes negative if $\psi'(l)\neq 0$.
So the redshift function is independent of $l$, an is chosen to be unity. For unit redshift function and shape
function of the form Eq. (16), the perturbation equation for the metric potential Eq. (12) takes the form

\begin{equation}\label{eq18}
	(\delta r(l))'' +{\frac {2(\delta r(l))'}{\omega L }}+\frac{\omega -1}{\omega^2 L^2}\delta r(l)=0.
\end{equation}

\begin{figure*}[thbp] \label{Rad2}
	\centering
	\includegraphics[width=0.5\textwidth]{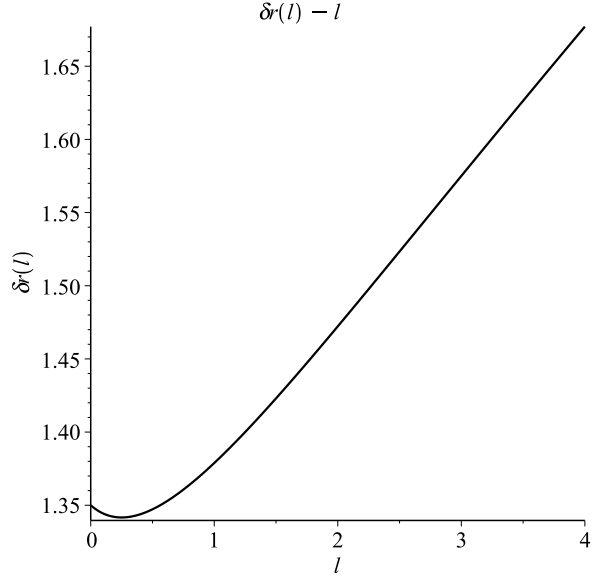}
	\caption{Variation of the perturbation with respect to $l$.}
\end{figure*}

The solution to this differential equation is given as

\begin{equation}\label{eq19}
	\delta r(l) ={K_3}( L)^{-\frac{1}{\omega}}+{K_4} (L)^{{\frac {\omega-1}{\omega}}},
\end{equation}
where $K_3$ and $K_4$ are the constants of integration.
We have plotted the variation of $\delta r$ with respect to the radial distance $l$ in Fig. 6. From the figure and from Eq. (\ref{eq19}) it can be found that the perturbation remains finite for all proper radial distance, hence the stability of the wormhole can be confirmed.

\subsubsection{NEC}

For the tachyon potential Eq. (17) obtained corresponding to the shape function Eq. (16) we check the validity of the NEC.
The term $\rho+ p_r$ can be evaluated as

\begin{equation}\label{eq20}
	\rho+p_r=-{\frac {l^4 \sqrt {2{\omega}^{2}{r_0}^{2} \left( \omega-2 \right) L^{{\frac {2\omega+2}{\omega}}}- \left( 2\omega-3 \right) {r_0}^{4} L^{4{\omega}^{-1}}+{\omega}^{4} L^{4}}}{\sqrt {{C_1} \left( -l^4+{C_1} \right) }{r_0}^{2}{\omega}^{2}} L ^{{\frac {-2-2\omega}{\omega}}}}.
\end{equation}

Here $L=1+l$. The term $\rho+p_r$ has been plotted against the proper radial distance $l$. Both from the
expression and figure we see that the NEC is violated. Hence the exotic tachyon matter prevents the throat from collapsing.

\begin{figure*}[thbp] \label{NEC2}
	\centering
	\includegraphics[width=0.5\textwidth]{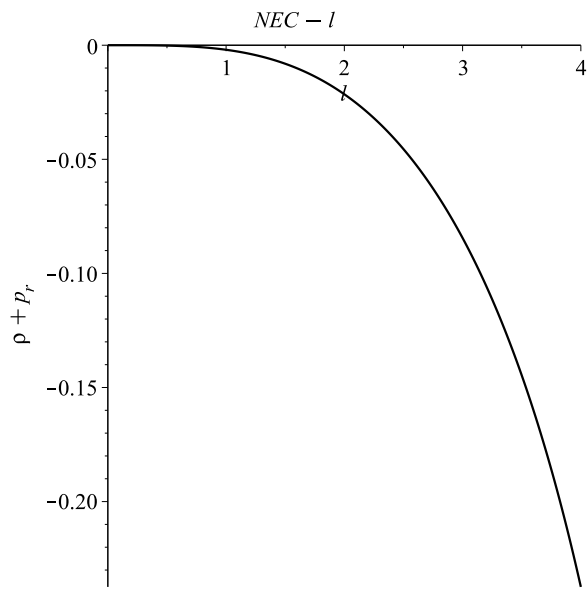}
	\caption{Variation of the NEC with respect to $l$.}
\end{figure*}

\section{Third shape function}

We consider another shape function of the form~\cite{Chakraborty}

\begin{equation}\label{eq21}
	r(l)=lt+r_0.
\end{equation}

Here $t$ is a parameter such that $0<t<1$.

\begin{figure*}[thbp] \label{VLT3}
	\centering
	\includegraphics[width=0.5\textwidth]{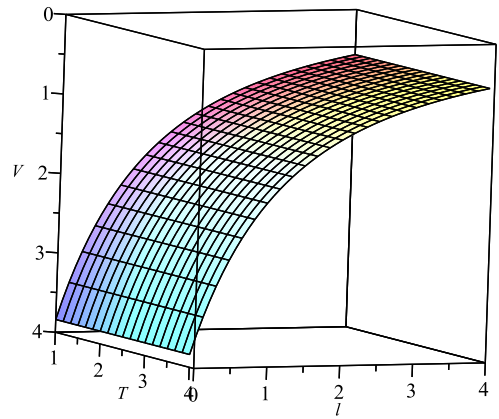}
	\caption{Variation of the V and T with respect to $l$.}
\end{figure*}

Once this form of the shape function $r(l)$ and the tachyon field $T(l)$ are used in the first EFE, we get the
tachyon potential as a function of the proper radial distance. The potential is found to have the form
\begin{equation}\label{eq22}
	V(l)={\frac {-{t}^{2}+1}{ \left( lt+r_{{0}} \right) ^{2}}}.
\end{equation}

As for the previous two shape functions, in this case also the tachyon field is independent of it, and we cannot
express the tachyon potential in terms of the field. So we draw a 3 D graph to show the variation of $V,T$ and $l$ in Fig. 8.

\subsection{Stability}

We will compute and plot the perturbation in the metric potential $r$. For doing so we use the perturbation Eq. (12)
inserting the present shape function into which modifies the perturbation equation to

\begin{equation} \label{eq23}
	(\delta r(l))'' +{\frac {2 t{\frac {\rm d}{{\rm d}l}}\delta r(l) }{lt+r_{{0}}}}=0.
\end{equation}

\begin{figure*}[thbp] 
	\centering
	\includegraphics[width=0.5\textwidth]{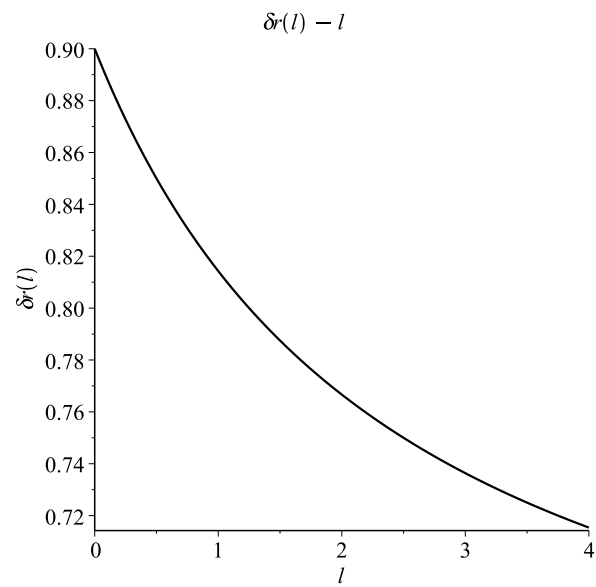}
	\caption{Variation of the radial perturbation with respect to $r$.}
\end{figure*}

The above differential equation can be solved easily to obtain an expression for the perturbation as

\begin{equation}
	\delta r(l) ={H_1}+{{H_2} \left( l+{\frac {r_{{0}}}{t}} \right) ^{-1}},
\end{equation}
where ${H_1}$ and ${H_2}$ are the constants of integration.
The perturbation is plotted along the proper radial distance in Fig. 9. As the perturbation can be seen to be
finite everywhere, we confirm the stability.

\subsubsection{NEC}

For this shape function also it is essential to check whether the obtained potential for the tachyon field is such that it
violates the NEC. So, we compute, as for the previous shape functions

\begin{equation}
	\rho+p_r=-{\frac { \left( -{t}^{2}+1 \right) l^4}{(lt+r_0)^{2} \left(-l^4+C_1\right)}{\frac{1}{\sqrt {1+{\frac{l^4}{-l^4+C_1}}}}}}.
\end{equation}

\begin{figure*}[thbp] \label{NEC3}
	\centering
	\includegraphics[width=0.5\textwidth]{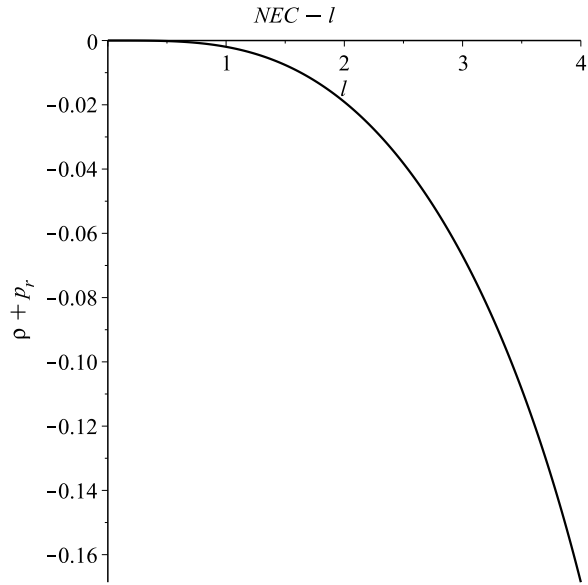}
	\caption{Variation of the NEC with respect to $l$.}
\end{figure*}

A plot of $\rho+p_r$ along the proper radial distance is given in Fig. 10. As we can see, for wormhole with this shape function supported by tachyon matter, the NEC is violated.

\section{Discussion and Conclusion}

In GR as well as theories of modified gravity, wormholes are a set of solutions to the EFE just
like black holes and gravitational waves. The question of their traversability and whether they allow the existence of closed time-like curves facilitating both
backward and forward travel in time, is a question that requires further investigation. Wormholes composed of ordinary matter in the context
of GR are found to be unstable as the structure pinches off at the throat region due to gravitational collapse. However,
as suggested by Morris and Thorne\cite{MT1988} three decades ago, if the throat contains matter which violates the NEC, then the pinch-off
can be prevented making the wormhole traversable. Such matter is called exotic matter. However, if we are to consider wormholes as solutions to the modified Einstein Field Equations (EFEs) in theories of modified gravity, then there is ususally no need for the wormhole to be supported by exotic matter as the modified geometry provides an ``effective matter" description which causes the effective energy density and pressures to violate the NEC~\cite{Sengupta2,Eiroa,Bertolami,Moraes,Agnese,He,Dzhunushaliev,Bronnikov,Lobo2,Rahaman1}.

So, in order to obtain meaningful and traversable wormhole solutions of physical interest, it is essential to satisfy the flare-out condition at the throat which ensures that the NEC must be violated either by introducing exotic matter at the throat or modifying the gravity framework leading to modified field equations which allow violation of NEC with non exotic matter. This picture is very similar to the one involving the physics of the ``\textbf{dark universe}." In order to explain the presence of dark matter and dark energy, if we keep ourselves confined to GR, then we must introduce substantial modification to the matter-energy source. The dark matter candidates are often known to be of the exotic form~\cite{Moskowitz,Sivaram}, and dark energy also leads to the violation of some of the energy conditions~\cite{Visser,Lasukov,Schuecker,Wu}. However, in this case also, if we switch to modifying the gravitational action and in turn the field equations, then the introduction of exotic matter is no longer a necessary requirement~\cite{Faraggi,Maia,Vasak,Chak2}. This is a striking similarity between the physics of the ``\textbf{dark sector}" and wormhole physics, hinting at the possibility of a deeper connection between the two. It may vey well be that the driving component of the late-time acceleration of the universe is the supporting matter for wormholes present at the throat and also the same component may form a fraction of the missing ``dark" matter composition, just like tachyon matter in our case.

In our work, we have considered the exotic matter in the form of tachyon matter. Das and Kar~\cite{Das} have constructed Ellis
wormhole using tachyon matter. We consider a number of other wormhole shape functions and check whether a stable wormhole can
be constructed which violates the NEC. We also want to obtain the tachyon field and the corresponding field potential. For the
Ellis wormhole~\cite{Das} had found that for realistic solutions of the tachyon field the redshift function must be a constant.
For simplicity, they chose the constant to be unity without loss of generality. From our analysis, we find that not only for the
Ellis shape function but for other shape functions representing a wormhole, in order for the tachyon field to be real, the redshift
function must be a constant which we also choose to be of unit order of magnitude.

Another essential feature of their analysis was that in the absence of the \textbf{cosmological constant} $\Lambda$, the obtained potential
turned out to be imaginary along with a constant tachyon field. However, we find that this property is peculiar to the Ellis wormhole,
unlike the property of the constant redshift function which is general for a tachyon matter supported wormhole. For other shape
functions considered by us in our analysis, even in the absence of a $\Lambda$-term in the EFE, we have obtained the potential to
be real and the tachyon field is an Elliptic function of the proper radial distance, which however is independent of the shape
function for generalized wormhole geometry and has a kink-like nature identical to the one obtained by them for the Ellis wormhole.

From the three dimensional plots for $V$, $l$ and $T$ we can observe that as desired for the effective four dimensional, low-energy
behavior for tachyon condensate matter, at the origin of the field corresponding to $T=0$ for our case, the potential $V$ exhibits
a maximum for all the three shape functions. This can be better visible if we plot the potentials and field separately along the
proper radial distance in two different 2-dimensional plots, but it can also be stated from the 3D plots as well. Also, the kink-like
configuration of the tachyon field obtained and the violation of the NEC for the potentials obtained corresponding
to all the three geometries justify that the nature of the solutions obtained are ideal for the construction of wormhole geometry.

We have also checked the linear stability of the different geometries against small perturbations. The first order perturbation in
the metric potential has been obtained in each case from a condition evolving out of the field equations. From the expressions of
the perturbation obtained for all three different geometries and also from their plots along the proper radial distance, the stability
of the wormhole geometries have been confirmed. For the first two geometries, the perturbation in the metric takes a monotonic increment
after a slight initial drop, while for the third shape function there is a monotonic decrease in the small perturbation. The perturbation
remains finite for all values of proper radial distance.

The possibilities of detection of wormholes have been studied in~\cite{Li,Ohgami,Tsukamoto,Nandi,Shaikh2}. A possible way of observing a wormhole may arise from the idea of the flux being not conserved separately in the spacetimes joined by the wormhole structure ~\cite{DS}. The idea may find application in studying this effect of wormhole on the orbit of stars near the black hole at our galactic centre. The lensing effect due to wormhole, identical to gamma ray bursts can suggest a possible upper limit on mass of wormhole ~\cite{Torres}. There may be a possibility of radiation pulse emission from wormhole which may be a possible mode of detection~\cite{D}. The investigation of scattering properties around a rotating traversable wormhole can differentiate it from a black hole either by exhibiting super radiance or non-identical quasinormal ringing at dominant multipoles~\cite{KZ}. Also, there may be existence of background quasinormal modes with a long lifetime for wormhole having variable redshift function~\cite{C}.

Thus, we can conclude that tachyon matter can be used as form of exotic matter for constructing wormhole not only with Ellis geometry,
but other well defined wormhole geometries as well. However, there is a constraint on the tachyon wormhole in the sense that for the
wormhole geometries studied with tachyon matter, the redshift function must be a constant but as we have shown clearly that for geometries other
than Ellis wormhole, there is no need to include an additional cosmological constant term in the field equations in order to obtain
physically plausible solutions. \textbf{Even for wormhole geometries without exotic matter in the modified gravity context, the redshift function is sometimes taken to be a constant.} Just like it can be attempted to describe the late-time cosmic acceleration independently with the tachyon field, here we find that the tachyon matter can be used independently to construct traversable wormholes without the need for a $\Lambda$-term.

\end{document}